\documentclass[a4paper,english,preprint,aps,pra]{revtex4-1}
\usepackage[T1]{fontenc}
\usepackage[utf8]{inputenc}
\usepackage{amsmath}
\usepackage{amssymb}
\usepackage{esint}
\usepackage{graphicx}
\usepackage{xcolor}



\usepackage{babel}
\begin{document}
\title{
 Quantum interference  in strong-field ionization by a linearly polarized laser pulse, and its relevance to tunnel exit time and momentum
}

\author{Szabolcs Hack\textsuperscript{1}, 
Szilárd Majorosi\textsuperscript{2}, 
Mihály G. Benedict\textsuperscript{2}, 
Sándor Varró\textsuperscript{1,3}
and Attila Czirják\textsuperscript{1,2}}

\email{czirjak@physx.u-szeged.hu}

\affiliation{
 \textsuperscript{1}ELI-ALPS, ELI-HU Non-Profit Ltd.,   W. Sandner str. 3, H-6728 Szeged, Hungary 
 \\
\textsuperscript{2}Department of Theoretical Physics, University of Szeged, Tisza L. krt. 84-86, H-6720 Szeged, Hungary
\\
 \textsuperscript{3}Wigner Research Centre for Physics,  Konkoly-Thege M. str. 29-33, H-1121 Budapest, Hungary
}


\begin{abstract}
We investigate the liberation of an atomic electron by a linearly polarized single-cycle near-infrared laser pulse having a peak intensity that ensures tunneling. 
Based on  phase space analysis and energy distribution in the instantaneous potential, we reveal the importance of quantum interference between  tunneling and  over-the-barrier pathways of escape. 
Tunneling is blurred both in space and time, and the contribution of 
tunneling at the mean energy is almost negligible.  
We suggest and justify improved initial conditions for a classical particle approximation of strong-field ionization, based on the quantum momentum function, 
and we show how to reconstruct them from the detected momentum of an escaped electron. 
\end{abstract}

\maketitle



Strong-field ionization of atoms plays a fundamental role in attosecond
physics \citep{Krausz_Ivanov_2009_RMP,farkas_toth}: a suitably strong laser pulse enables
an electron to escape from its atomic bound
state into the continuum, usually assumed 
to happen by
tunneling, which is the first step of the very successful
three-step model underlying much of our understanding 
in this area
\cite{Corkum_PRL_1993,varro1993multiphotonequation,lewenstein1994hhgtheory,becker1994tdsehhgmodel,ivanov2005strongfield}. 
Currently, the problems of tunneling
time and exit momentum 
are of outstanding importance regarding both quantum theory and 
attosecond metrology 
 \citep{Hickstein_Kapteyn_PRL_2012_tunneling_distance,Guo_Becker_Agostini_DiMauro_PRL_2013_LES_ATI,Guenot_Balogh_Varju_Lindroth_LHuillier_Dahlstrom_JPhysB_2014_photoem_timedelay,Landsman_Keller_2015_PhysRep,Kheifets_JPB_2020}.
Several research groups published relevant experimental results
\citep{Comtois_Corkum_2005_JPhysB,
Eckle_Science_2008,
Schultze_Science_2010,
Pfeiffer_Keller_2012_PRL,
shafir_nature_2012,
Hofmann_Keller_PRA_2014,
Wolter_Moshammer_Biegert_PRX_2015,
Pedatzur_Dudovich_2015_NatPhys,
Camus_Moshammer_2017_PRL,
Porat_Krueger_Dudovich_NatComm_2018,
Sainadh_Litvinyuk_Nature_2019,
Li_Lu_Wuhan_PRL_2019_interferometry_LongitudMom_exp},
and some of these indicate a non-zero longitudinal exit momentum  
\citep{Comtois_Corkum_2005_JPhysB,
Pfeiffer_Keller_2012_PRL,
Hofmann_Keller_PRA_2014,
Wolter_Moshammer_Biegert_PRX_2015,
Pedatzur_Dudovich_2015_NatPhys,
Camus_Moshammer_2017_PRL,
Li_Lu_Wuhan_PRL_2019_interferometry_LongitudMom_exp}, 
either based on the attoclock method
 which relies on nearly  circularly  polarized pulses \citep{Eckle_Science_2008,Pfeiffer_Keller_2012_PRL,Torlina2015},
 or using linearly polarized pulses, sometimes along with weak auxiliary fields.
 For 
most of the established methods to generate attosecond pulses \citep{hentschel2001attosecond,baltuska2003attosecond,sansone2006attosecondisolated,Liu_JO_2018_MPQ_attostreaking,Ye_Varju_ELI_JPhysB_2020},  laser pulse polarization is linear (at least in the middle of the pulse) and 
the Keldysh-parameter $\gamma$  \citep{Keldysh_JETP_1965} is close to 1, i.e. it is beyond the validity range of well-known theories \citep{Delone_Krainov_1998}.
Recent theoretical approximations 
of strong-field ionization meeting such conditions
often employ classical dynamics, 
where the choice of proper initial conditions, including the longitudinal momentum, is an important open question  \citep{Yudin_Ivanov_2001_PRA,Eberly_PRL_2013_VirtDet,Hofmann_Keller_PRA_2014,Teeny_PRL_2016,
Ni_Rost_PRL_2016_classic_backprop_ell,Tian_Wang_Eberly_PRL_2017_num_detector_longmom,Douget_PRA_2018,Klaiber_PRL_2018,Xu_PRA_2018},
 with controversies regarding e.g. conservation of energy.

In this Letter, 
we reveal the real classical picture that can be associated with the exact quantum process of 
strong-field ionization
of a single atom driven by a linearly polarized few-cycle laser pulse, in the $\gamma \approx 1$ range. Our analysis using the Wigner function over the  phase
space inspires improved initial conditions for a set of classical electron trajectories which approximate the quantum evolution very well.
Regarding model parameters, we focus on atomic hydrogen driven by a few-cycle  pulse with a near-infrared carrier wavelength and a peak intensity 
in the 100  TW/cm$^2$ range,
which is a wide-spread case in theoretical works based on its relevance to state of the art experimental techniques.
We use atomic units unless otherwise stated.


We work in the framework of a simple model: we use dipole approximation for
the interaction of a single active electron atom with the classical
electromagnetic field in the length gauge. 
We define the laser pulse by the  temporal pulse shape of its
electric field along the $z$-direction 
for $0<t<NT$ 
as: 
${\cal E}_{z}(t)=F\cdot\sin^{2}\left({\pi t / N T}\right)\cos\left({2\pi t / T}+\phi \right)$,
having $N$ optical cycles of period $T$ and a carrier-envelope phase difference (CEP) of $\phi$. This pulse excites the electron from its atomic ground state. The electron's wave function then does not depend on the azimuth angle around the $z$ axis: $\Psi = \Psi \left(z,\rho,t\right)$  thus 
we can write the three-dimensional (3D) time-dependent
Schrödinger equation  (TDSE) for the electron's motion (assuming a fixed nucleus)  in cylindrical
coordinates $(z,\rho)$ as 
\begin{equation}
i\frac{\partial}{\partial t} \Psi =  
\bigg[  -\frac{1}{2}\left(\frac{\partial^{2}}{\partial z^{2}}+\frac{\partial^{2}}{\partial\rho^{2}}+\frac{1}{\rho}\frac{\partial}{\partial\rho}\right) 
+ \, V(z,\rho,t) \bigg] \Psi, 
\label{eq:3DTDSE}
\end{equation}
with 
$V(z,\rho,t)=-{1}/{\sqrt{z^2+\rho^2}}  +{\cal E}_{z}(t)\, z$.
Our recently developed algorithm \cite{majorosi2016tdsesolve} supports the direct numerical
integration of this TDSE with Coulomb-singularities and provides fourth
order accuracy in both space and time. 
In the following, the numerical results assume a laser pulse with $N=3$,  $T=110$ (corresponding to ca. 800 nm carrier wavelength) and $\phi=0$,
having an electric field amplitude of $F=0.06$ (corresponding to ca. $1.26 \cdot 10^{14}$ W/cm$^2$ peak intensity). 
These ensure that 
tunneling is possible during the entire laser pulse, $\gamma = 0.952$, and the pulse length is short enough to model high-order harmonic generation (HHG) resulting an isolated attosecond pulse \cite{Christov1997,hentschel2001attosecond,baltuska2003attosecond}.

Since the main  dynamics happens along the 
laser polarization \citep{majorosi2018densitymodel}, 
and a phase space analysis is more feasible in one dimension (1D),
we  compute the 1D reduced density matrix of the 
quantum state along the $z$ axis from 
$\Psi$ by integrating over the azimuth angle and radial coordinate:
\begin{equation}
\varrho (z,z',t) = 2 \pi \int_{0}^{\infty} \Psi^{*} \left(z,\rho,t\right) \Psi \left(z',\rho,t\right) \rho \, d\rho .
\end{equation} 
Then we compute the corresponding Wigner function \citep{BOOK_QUANTUM_OPTICS_2011} 
\begin{equation}
W(z,p_z,t)=\frac{1}{\pi}\int_{-\infty}^{\infty} \varrho\left(z+\zeta,z-\zeta,t\right)  {e}^{2 i p_z\zeta} \, d\zeta,
\label{eq:Wig_def}
\end{equation}
which is a successfully used tool also in strong-field and attosecond physics  \citep{czirjak2000ionizationwigner,graefe2012quantumphasespace,Schleich_Varro_PhysLettA_2013_tunnel_parabolic_Wigner,baumann2015strongfieldwigner,CarlaMorissonFaria_NJP_2019,Han_Liu_Beijing_PRL_2019_improved_attoclock_Wigner}.
%
The main advantage of an analysis based on the Wigner function is that $W$ displays the quantum description in a close analogy to the classical phase space dynamics, thus it enables to use our intuition based on classical physics,  
while it still contains all the quantum details, most notably quantum interference
($W$ is equivalent to $\varrho$, 
since Eq. (\ref{eq:Wig_def}) can be inverted).
\begin{figure*}[ht]
\includegraphics[width=0.353\columnwidth]{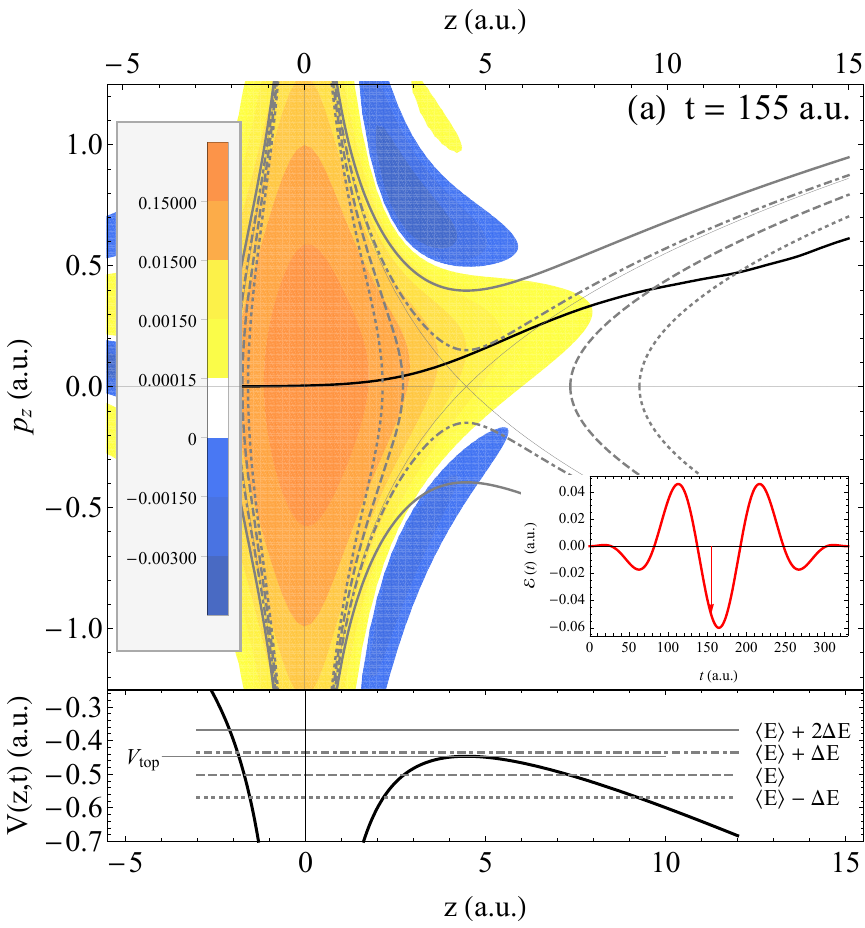}
\includegraphics[width=0.312\columnwidth]{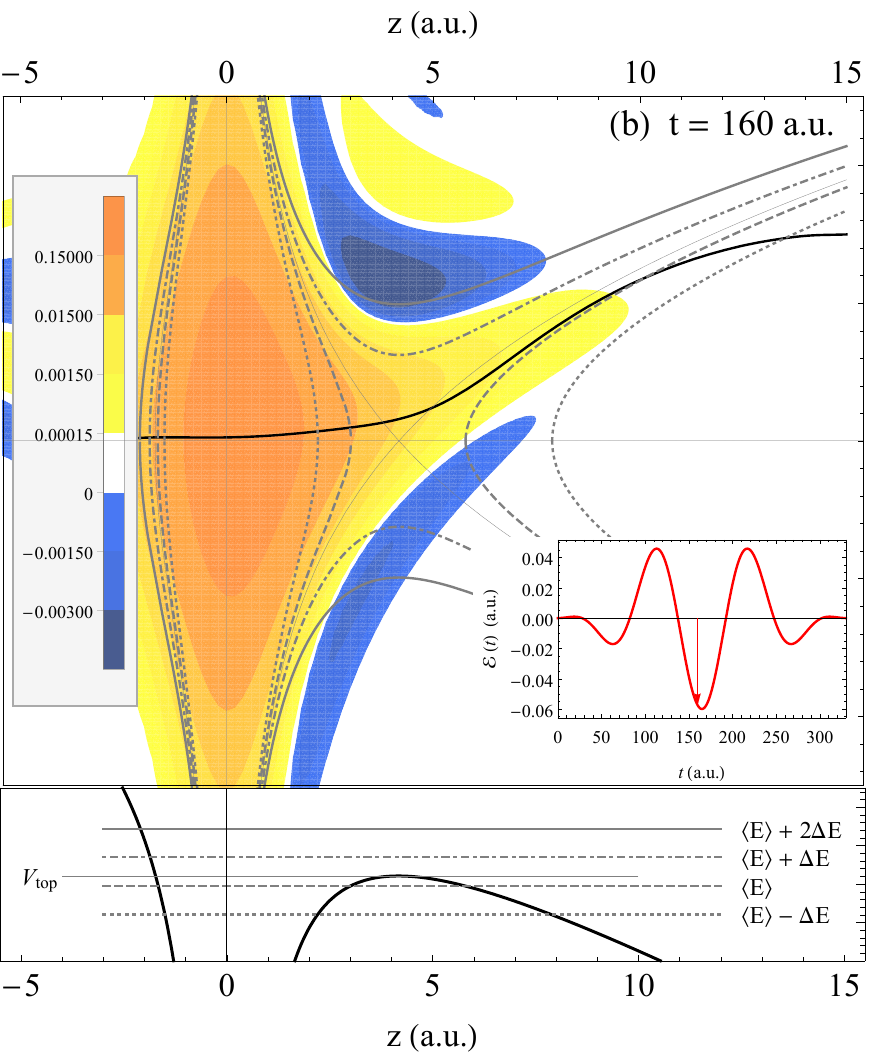}
\includegraphics[width=0.312\columnwidth]{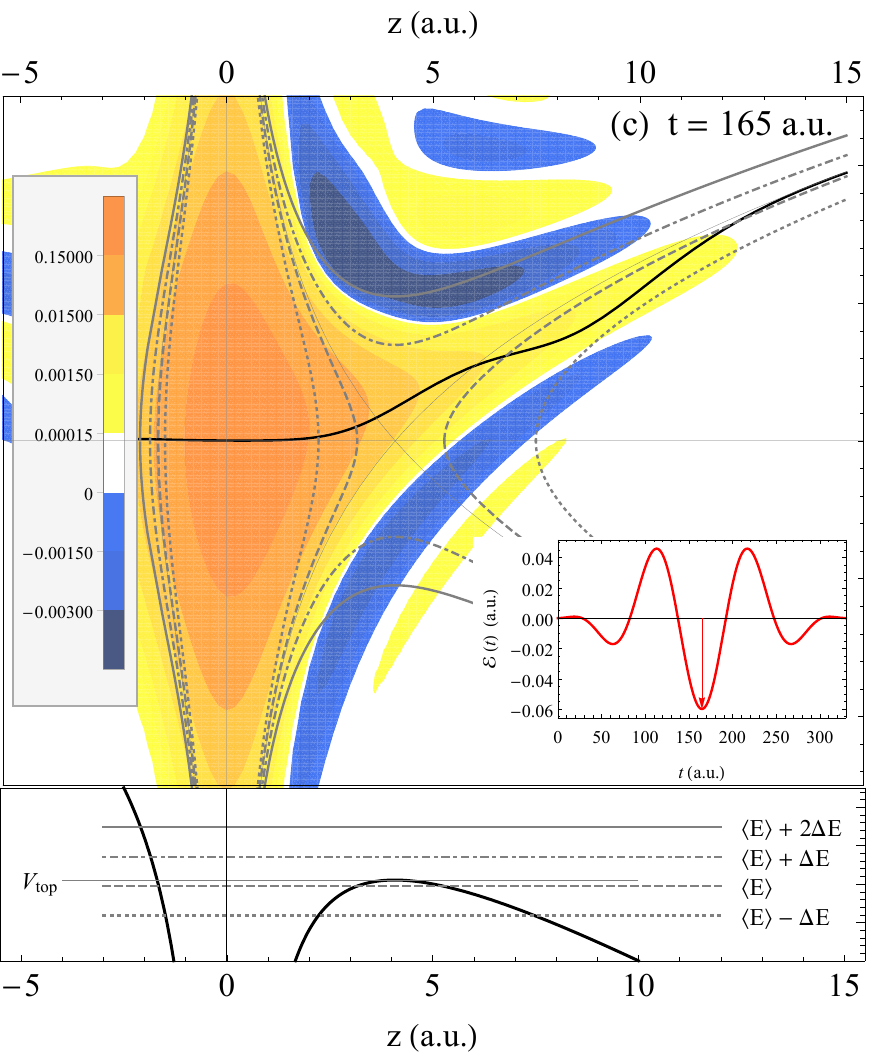}

\protect\caption{Contour plots of the reduced 1D Wigner function of Eq. (\ref{eq:Wig_def}) in logarithmic color scheme (upper panels) and plots of $V(z,0,t)$ (lower panels, in black) at $t=155$ (a), $t=160$ (b) and $t=165$ (c) during the liberation of the hydrogen atom's electron by the laser pulse of ${\cal E}_{z}(t)$ plot in the insets. 
Phase space trajectories in gray with different styles represent classical motion in the instantaneous potential, corresponding to energy levels marked in the lower panels with respective styles. The trajectory with thin gray solid line is the instantaneous separatrix between tunneling and OB ionization regimes.
Black solid lines in the upper panels show the quantum momentum function of Eq. (\ref{eq:qm}).}
\label{fig:Wigner_and_trajectories} 
\end{figure*}

\begin{figure}[h]
\includegraphics[width=1\columnwidth]{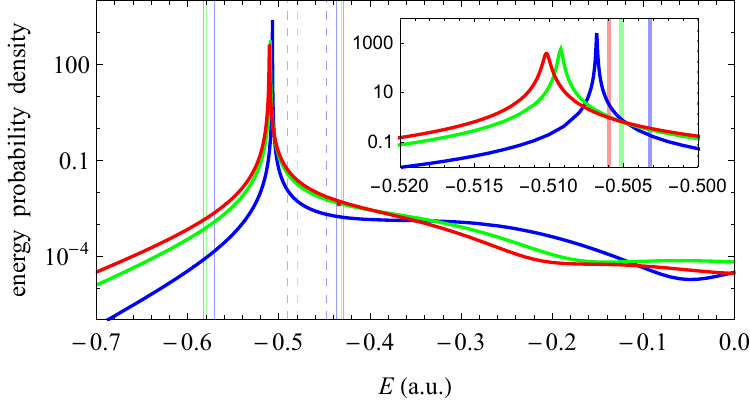}
\protect\caption{Energy probability density of the quantum state at $t=155$ (blue), 160 (green) and 165 (red) in the instantaneous potential. Thin solid vertical gridlines mark the $\langle E \rangle (t) \pm \Delta E (t)$ values, dashed gridlines mark $V_\mathrm{top}(t)$, in respective colors. The inset zooms into the region around the peaks, thick solid gridlines mark $\langle E \rangle (t)$.}
\label{fig:energy_distribution}
\end{figure}

Snapshots of this Wigner function at selected time instants close to and at the peak intensity of the laser pulse 
in Fig. \ref{fig:Wigner_and_trajectories} 
clearly show a developing stream (in yellow) 
along classical phase space trajectories of escaping particles, 
which already suggests that the liberation of the electron is blurred in time. 
 The oscillating ripples, including regions with negative function values, refer to strong quantum interference, similarly to earlier results with a  simpler model \citep{czirjak2000ionizationwigner}.
The Wigner function's contour lines well follow the stationary phase space trajectories of classical particles with relevant energies in the instantaneous potential 
$V(z,\rho,t)$, including over-the-barrier (OB) ionization paths.
This feature suggests the analysis of the energy distribution of the momentary quantum state, which we performed numerically
in the instantaneous eigenstate basis of $V$,
 since we think this more appropriate then to follow the population of the  unperturbed atomic bound states \citep{Serebryannikov_Zheltikov_PRL_2016_tunneling_excited_bound_states_theo}.
The instantaneous energy probability densities, shown in Fig. \ref{fig:energy_distribution}, 
are sharply peaked around energy values which are close to but lower than the mean energy $\langle E \rangle (t)$. 
These peaks get broader and lower, and the  population of the energy range above the top of the potential barrier $V_\mathrm{top}(t)$ increases considerably, as the laser pulse approaches its peak. Thus, the energy variance $\Delta E (t)$ increases as well.  

\begin{figure}[h]
\includegraphics[width=1\columnwidth]{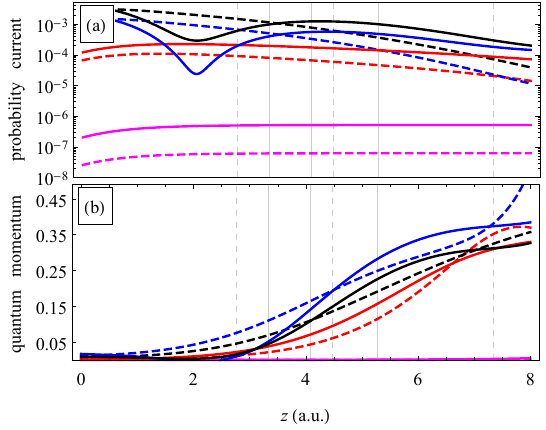}
\protect\caption{Comparison of the probability currents $j(z,t)$ (black),  $j_{\mathrm{FT}}(z,t)$ (red), $j_{\mathrm{OB}}(z,t)$ (blue) and  $j_{\mathrm{ST}}(z,t)$ (magenta) in (a), and the corresponding quantum momentum functions in (b). We plot these curves at the peak of the laser pulse $t=165$ (solid lines) and somewhat earlier at $t=155$ (dashed lines). The vertical gridlines mark the momentary positions of the  tunnel entrance, the $V_\mathrm{top}(t)$ and the tunnel exit, from left to right, with respective style.}
\label{fig:prob_current_and_qm_plot}
\end{figure}

In order to evaluate the importance of the tunneling and OB ionization pathways, we compare the  1D probability currents $j_{\mathrm{FT}}(z,t)$ and $j_{\mathrm{OB}}(z,t)$ of the 
respective 
full tunneling (FT) and OB
wave packets, shown in Fig. \ref{fig:prob_current_and_qm_plot}(a), which we compute using the instantaneous energy eigenstate decomposition, by integrating over all of the  corresponding energy range below and above $V_\mathrm{top}(t)$, respectively. These curves suggest that the resulting coherently summed full OB pathway is at least as important as the resulting coherently summed full tunneling pathway of ionization. Comparing them to the total  probability current $j(z,t)$  obviously shows strong interference between them. 
We would like to emphasize that this full tunneling pathway significantly expands  
the most widespread view of tunnel ionization, when tunneling is considered usually at the initial bound state energy  or at  the  mean value of the energy. 
In order to highlight this difference, we also plot the probability current $j_{\mathrm{ST}}(z,t)$ of a such a sharp tunneling
(ST)
 wave packet, which we create by integrating over a narrow energy range of 0.004 a.u. centered at the mean value of the energy. The  $j_{\mathrm{ST}}(z,t)$  is 2-3 orders of magnitude smaller than the full tunneling current  $j_{\mathrm{FT}}(z,t)$, and this fact does not change qualitatively, if the narrow energy range is centered at the peak of the energy probability density. 
This seemingly surprising difference may be well explained by the "laser acceleration" beyond the tunnel exit for those components which tunneled out at higher energy levels. 
However, the above results actually question the justification of a sharply defined position and time instant for the tunnel exit.

Due to the essential role of quantum interference during the escape, 
the classical particle model of the escaping electron should account for all of the possible pathways.
One of the most important open questions for such a classical particle is  a properly chosen initial momentum.
This is in close connection with the Wigner function,
since its meaning leads intuitively  to the notion of a position-dependent average momentum in terms of its $n$th moment  
$P_{n}\left(z,t\right)=\int_{-\infty}^{\infty} p_z^{n}\,W(z,p_z,t)\,dp_z$ as
\begin{equation}
q\left(z,t\right)=
P_{1}\left(z,t\right)/P_{0}\left(z,t\right),
\label{eq:qm}
\end{equation}
which is in fact identical to the gradient of the Madelung phase (multiplied by $\hbar$),
usually called quantum momentum function or flow momentum \citep{Madelung_1927}. 
We plot the  quantum momentum function $q(z,t) $ in the phase space snapshots of Fig. \ref{fig:Wigner_and_trajectories} with a black solid line: 
it follows very well   the main stream 
of the Wigner function corresponding to an escaping particle in the spatial domain where the electron's probability density is significant. 
Its developing oscillations are due to quantum interference of the ionization pathways with different energies.
Fig. \ref{fig:prob_current_and_qm_plot}(b) shows clearly the interference between the quantum momenta of the FT and OB wave packets, 
and the negligible quantum momentum of the ST wave packet.

\begin{figure}[ht!]
\includegraphics[width=1\columnwidth]{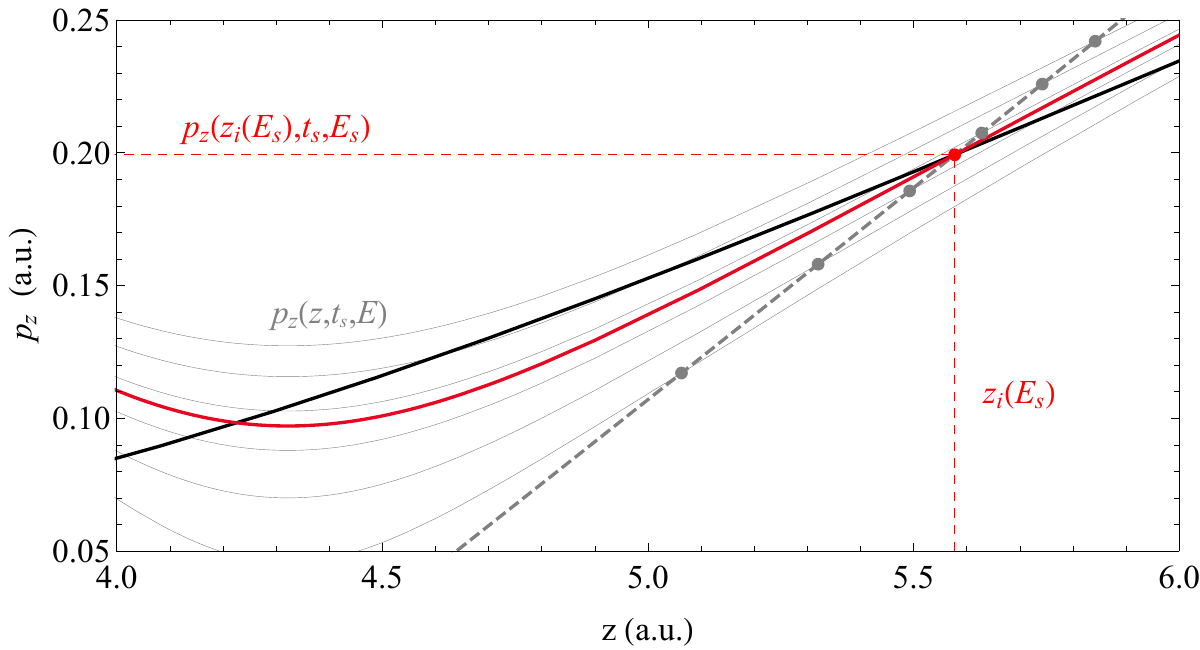}
\protect\caption{
Definition of the initial phase space point 
(shown for $ t_{\mathrm{s}}=157$): there is only a single inflection point (in red) among the inflection points (dashed line and dots in gray) of the stationary phase space trajectories  (gray solid lines) of the instantaneous potential,  which is on the instantaneous quantum momentum function (black solid line).
}
\label{fig:init_cond}
\end{figure}

Next we show that properly chosen initial conditions enable a suitable set of classical trajectories that follow the quantum momentum function very well  
and they also reflect that the liberation  process is blurred both in space and time. 
Based on its physical meaning, 
it is intuitive to use a suitable point of  the quantum momentum function itself as the initial condition of a possible classical trajectory 
which starts at the instant $t_{\mathrm{s}}$ from the phase space point $\left( z_{\mathrm{s}}, q(z_{\mathrm{s}},t_{\mathrm{s}}) \right)$. 
This initial $z_{\mathrm{s}}$ coordinate should be chosen in such a way, that 
the quantum propagation of the escaping wave packet beyond  $z_{\mathrm{s}}$ can be  approximated by classical dynamics  with sufficient accuracy,
but it is still close enough to the position of the local maximum of the potential barrier so that it is relevant as 
 initial position of an escaping trajectory.
A reasonable balance of the latter two requirements is provided by
 the position  $z_{\mathrm{i}}(E)$ of the outermost inflection point of
 a suitable stationary phase space trajectory of the instantaneous potential,
given by $p_z \left(z,t_{\mathrm{s}};E\right)=\sqrt{2\left(E-V\left(z,0,t_{\mathrm{s}} \right)\right)}$
for total energy $E$.
We plot these trajectories  and the corresponding inflection points  for a few  of the continuum values of $E$ in Fig. \ref{fig:init_cond}.
The $p_z \left(z,t_{\mathrm{s}};E\right)$ are concave for $z>z_{\mathrm{i}}(E)$, because the dynamics is dominated by the laser electric field, which allows of  a classical approximation.
Taking into account now all of the above considerations, it follows that the  initial position $z_{\mathrm{s}}$ we seek is
actually the position of that particular inflection point 
where  the corresponding stationary phase space trajectory 
intersects the quantum momentum function, see the red and black curves in Fig. \ref{fig:init_cond}.
Denoting the corresponding energy by $E_{\mathrm{s}}$, the $z_{\mathrm{i}}(E_{\mathrm{s}})$ selected by the quantum momentum function is
the solution of the following equation:
 \begin{equation}
q(z_{\mathrm{i}}(E_{\mathrm{s}}),t_{\mathrm{s}})=p_z \left(z_{\mathrm{i}}(E_{\mathrm{s}}), t_{\mathrm{s}}; E_{\mathrm{s}} \right).
\label{eq:new_ini_cond}
\end{equation}
At every realistic starting time $ t_{\mathrm{s}} $,  there is only a single solution $z_{\mathrm{i}}(E_{\mathrm{s}})$ 
of Eq. (\ref{eq:new_ini_cond}).
We propose this as the initial position, $z_{\mathrm{s}}=z_{\mathrm{i}}(E_{\mathrm{s}})$ and 
the unique  phase space point $\left( z_{\mathrm{s}}, q(z_{\mathrm{s}},t_{\mathrm{s}})\right)$ 
as the  initial state of the classical escape trajectory which starts at $t_{\mathrm{s}}$. 
This initial state is always in the OB regime but,
as we shall see, it is still a good classical representation of all possible escape pathways depending on the starting time.
\begin{figure}[ht!]
\includegraphics[width=1\columnwidth]{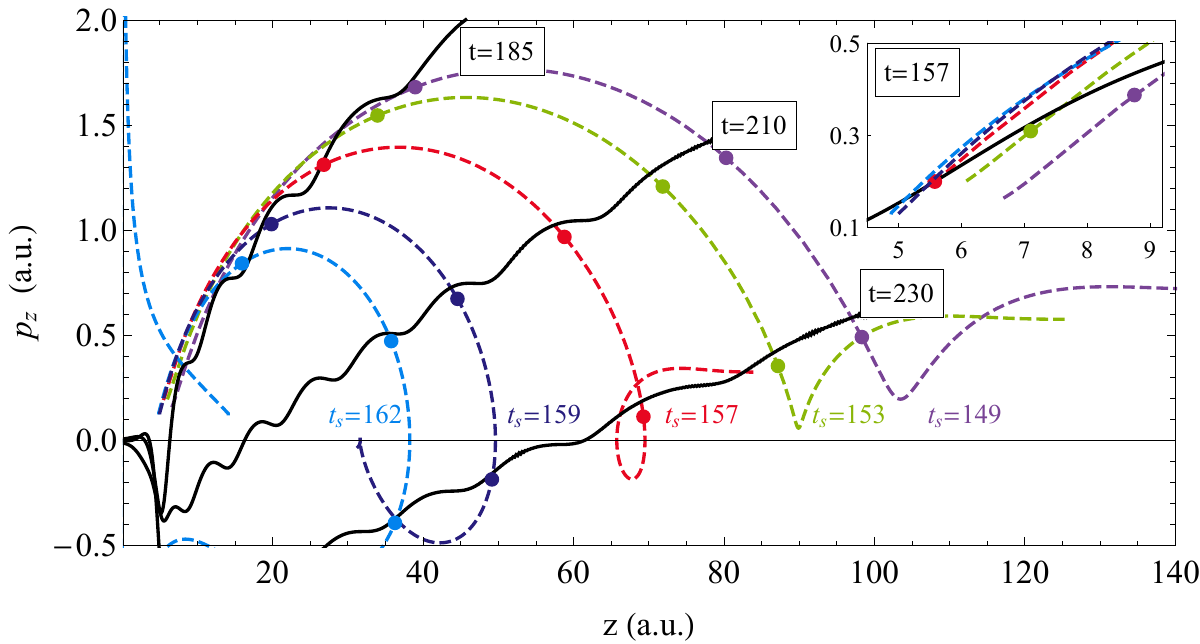}
\protect\caption{Classical phase space trajectories of the liberated electron (dashed lines) for the indicated starting time instants, plot from $ t_{\mathrm{s}} $ to $t=NT$. 
 Initial conditions correspond to Eq. (\ref{eq:new_ini_cond}), the inset zooms into the phase space region around these start points.
Directly escaping trajectories (in purple, green and red) are represented by $ t_{\mathrm{s}}=149,\,153,\,157$. Dark blue indicates a limiting case for $ t_{\mathrm{s}}=159 $ having ca. 0 momentum at $t=NT$. Light blue marks a rescattering trajectory $ \left( t_{\mathrm{s}}=165 \right) $.  
Snapshots of the classical and the quantum propagation at $ t=157,\,185,\,210,\,230$, represented by the colored dots and by $q(z,t)$ (black solid line), 
respectively, show a good match for all of the different outcomes.}
\label{fig:classical_trajectories}
\end{figure}
We plot several phase space trajectories and some characteristic snapshots of the dynamics of the escaping electron for different starting time instants in Fig. \ref{fig:classical_trajectories}. The electron is close to the quantum momentum function during the whole propagation, regardless of the starting time, which  well justifies that they represent a good classical approximation of the true quantum dynamics. As the $t_{\mathrm{s}}$ values get closer to the instant of the main peak of the laser pulse, for $t_{\mathrm{s}}>157$ the electron does not have enough kinetic energy at the end of the interaction 
to leave its parent ion  permanently. For further increasing starting time instants, the electron is  re-scattered with energy gain from the laser pulse and it may thus contribute e.g. to HHG. The starting positions are in good agreement with earlier data derived from experiments \cite{Hickstein_Kapteyn_PRL_2012_tunneling_distance}.

We can gain more insight into the  onset of the escape process by studying a wave packet which is defined to have 
positive momentum and positive energy (PMPE) at the final instant of the laser pulse: i.e. we obtain this PMPE wave packet by first subtracting the contributions of the bound states $ \left| n,\ell \right\rangle$  from the 3D solution at $t=NT$:
 \begin{equation}
\left|  \Psi_{\mathrm{PE}} (t=NT) \right\rangle = \left| \Psi (t=NT) \right\rangle - \sum_{n,\ell} \langle n,\ell | \Psi \rangle \left| n,\ell \right\rangle
 \end{equation}
 and then keeping (by Fourier-filtering) only those components of $\left|  \Psi_{\mathrm{PE}} (t=NT) \right\rangle$  which have positive $p_z$.
In an ideal case, this PMPE wave packet is able to reach a detector at a macroscopic distance (placed at a position with a large positive $z$ coordinate).
However, we now propagate this 3D PMPE wave packet backwards in time using the 3D TDSE  (\ref{eq:3DTDSE}) and then we do a phase space analysis of the resulting 
 1D Wigner function $W_{\mathrm{PMPE}}(z,p_z,t)$ which we obtain from the 3D PMPE wave packet just as we obtained $W(z,p_z,t)$ from  the full 3D wave function  $  \Psi\left(z,\rho,t\right)$. 
\begin{figure}
 \includegraphics[width=1\columnwidth]{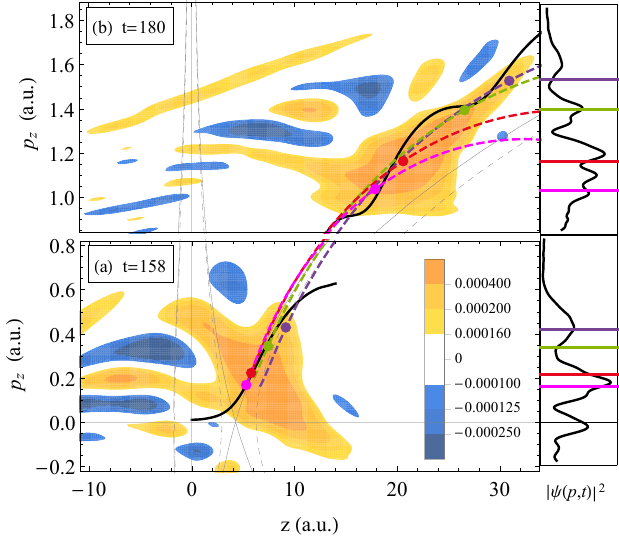}
\protect\caption{Comparison of classical and PMPE quantum dynamics of an escaping electron at $ t=158$ (a) and at $ t=180$ (b).  
Left panels: colored contour plots of $W_{\mathrm{PMPE}}$ and plot of $q(z,t)$ (solid black line). 
Solid and dashed thin gray trajectories are as in Fig. \ref{fig:Wigner_and_trajectories}.
Classical trajectories in purple, green, red and magenta dashed lines belong to  $ t_{\mathrm{s}}=149,\,153,\,157 $ and $ 158 $, respectively, having different initial conditions according to Eq. (\ref{eq:new_ini_cond}).
Dots in corresponding colors mark the momentary state of the classical dynamics.
Right panels:  momentum distribution of the PMPE wave packet, momentum values relevant to classical dynamics marked in respective colors. }
\label{fig:wigner_pmpe}
\end{figure}
We compare the  classical trajectories according to the initial conditions of Eq. (\ref{eq:new_ini_cond})
 with the $W_{\mathrm{PMPE}}(z,p_z,t)$ in Fig. \ref{fig:wigner_pmpe}. The snapshot at $ t=159 $ clearly shows that besides tunneling, 
 the OB pathway is essential also in the escape of the PMPE wave packet. 
 If a single "privileged" classical trajectory should be chosen, the one starting at $t=157$ a.u. seems to be the good representation of this process.
 However, a suitable set of such trajectories with different starting times gives apparently a better approximation of the quantum dynamics of the PMPE wave packet, and such a set reflects also the temporally blurred feature of the liberation process.
  This approach is further supported by the observation that the Wigner function approximately follows the classical propagation during its time evolution,
  in agreement with the well-known fact that for potentials with up to quadratic spatial dependence (which is a good approximation in the relevant spatial domain of $W_{\mathrm{PMPE}}$) the quantum Liouville equation is identical  to  the classical one.
  This makes the assignment of probabilities to these trajectories possible and meaningful.

Finally, we show how a simple procedure can reconstruct the starting time from the detected momentum of a directly escaped electron.
Neglecting the Coulomb interaction first, the momentum of the electron at the end of the laser pulse would be the following:
 \begin{equation}
p_{f}^{\mathrm{NC}}=p_0+\int_{t_{\mathrm{s}}}^{t_{f}} e\,{\cal E}_z (t)\,dt,
\label{eq:pfNC_ts}
 \end{equation}
where $ p_{0} $ is unknown if  $ t_{\mathrm{s}} $ or $q(z,t)$ is unknown. 
In reality, the escaping electron interacts with its parent ion's Coulomb potential until it reaches the detector at infinity where its measured momentum is
 $  p_{\mathrm{d}}=\sqrt{(p_{f}^{\mathrm{C}})^{2}+2\,V_{f}^{\mathrm{C}}}$
in terms of its potential energy $ V_{f}^{\mathrm{C}} $ and momentum
  $p_{f}^{\mathrm{C}} $ at $t_{f} =NT $.
Denoting by $ \Delta W $  the difference of the work of the laser electric field on the trajectories with and without the Coulomb interaction,
we get
 \begin{equation}
 p_{f}^{\mathrm{NC}}=\sqrt{ p_{\mathrm{d}}^{2}-2 \left( V_{0}^{\mathrm{C}}+\Delta W\right)},
 \label{eq:pd_pfNC}
 \end{equation}
where $ V_{0}^{\mathrm{C}} $ is the potential energy of the electron in the Coulomb field at $ t_{\mathrm{s}} $.
 The  energy in  parentheses can be approximated well in terms of the initial coordinate $z_0$ as  $V_{0}^{\mathrm{C}}+\Delta W = -1/z_{0}$. 
Eqs. (\ref{eq:pd_pfNC}), (\ref{eq:pfNC_ts}), and Eq. (\ref{eq:new_ini_cond}) with $q$ replaced by $ p_{0} $,  enable an iterative procedure to reconstruct the value of  $ t_{s} $ from $p_{\mathrm{d}}$, based on realistic starting values of  $ p_{0} $ and  $ z_{0} $. 
In order test the accuracy of this procedure, we generated the $p_{\mathrm{d}}$ values in a "numerical experiment": we computed the final momenta of escaping classical trajectories according to the initial conditions of Eq. (\ref{eq:new_ini_cond})  for a set of  $ t_{s} $ which are relevant for the PMPE wave packet. 
Then we applied the above iterative procedure to obtain the reconstructed values of the starting time: $ t_{s}^{R} $.
The results listed in Table  \ref{tab:starting_time} show that, although the largest error exceeds 2 a.u. for the earliest starting times, the accuracy greatly  improves for more probable trajectories and the error  may get even below 5 attoseconds for $ 155 \leq t_{s} \leq 157 $. Note that the corresponding phase space trajectories follow the larger values of the PMPE Wigner function. 

\begin{table}[h]
\caption{\label{tab:starting_time}Comparison of real staring times ($t_{s}$) and reconstructed starting times  ($ t_{s}^{R} $) obtained from the method described in the main text.}
\begin{ruledtabular}
\begin{tabular}{l*{7}{|l}}
$ t_{s} $  & 149 & 151  & 153 & 154 & 155 & 156 & 157\\
\hline
$ t_{s}^{R} $ & 151.88 & 153.11  & 153.99 & 154.47 & 155.06 & 155.84 & 156.84\\
\end{tabular}
\end{ruledtabular}
\end{table}

In conclusion, our results provide 
new insight into  the process
of atomic strong-field ionization by a linearly polarized single-cycle laser pulse.
We investigated the Keldysh parameter range of $\gamma \approx 1$, which is typical in many  of the relevant experiments but
it lacks exact analytic theory. Based on  accurate numerical simulations, 
important features of the electron's motion are shown by the Wigner function intuitively: the escaping wave packets create streams and oscillating ripples
as a manifestation of quantum interference,
which also rotate clockwise around the central bound part as the laser electric field drives the process.
Our primary purpose was to explore the quantum details of the electron's liberation around the main peak of the laser pulse.
 Interference between components tunneling with different energies (below $V_\mathrm{top}(t)$) 
 makes both the time and the position of the tunnel exit blurred, which questions the usual picture of strong field tunneling.
Both the Wigner function and  the probability currents show that the over-the-barrier  pathways  have an important 
contribution to the liberation, despite  $\langle E \rangle (t) < V_\mathrm{top}(t)$. 
These explain experimental results on non-zero longitudinal momentum at the tunnel exit, which apparently contradict energy conservation
according to the definition of a sharp tunnel exit.
We showed that the Wigner function  naturally inspires to use the quantum momentum function and 
 that this enables new initial conditions (Eq. (\ref{eq:new_ini_cond})) for classical approximation with non-zero initial longitudinal momentum. 
The resulting classical trajectories follow the quantum evolution very well, 
they account for direct escape and re-scattering, depending on start time, 
and a suitable set of such trajectories reflects  the  blurred nature of the liberation process.
The presented results were checked to be qualitatively valid for $F=0.04-0.06$ and $\phi=0, \, \pi/4,\, \pi/2,\, 3\pi/4$.
We believe that the Wigner function will be useful in further analysis of strong-field ionization, especially if quantum interference has an important role,
like e.g. in re-scattering, above threshold ionization, and low energy structures. 


\begin{acknowledgments}
We thank Wilhelm Becker, Péter Földi, Balázs Major and Katalin Varjú for valuable discussions.
This research was performed in the framework of the project Nr. GINOP-2.3.2-15-2016-00036,
and also supported by the European Union, co-financed by the European Social Fund, Grant
No. EFOP-3.6.2-16-2017-00005. 
Partial support by the ELI-ALPS project is also acknowledged. 
The ELI-ALPS project (GINOP-2.3.6-15-2015-00001) is supported 
by the European Union and co-financed by the European Regional Development Fund.
This work was supported by the Ministry of Innovation and Technology, Hungary grant NKFIH-1279-2/2020.
\end{acknowledgments}

\bibliographystyle{apsrev4-1}
\bibliography{revised_manuscript_Hack-Czirjak}

\end{document}